\documentclass[journal,10pt]{IEEEtran}

\usepackage[font=scriptsize,caption=false,labelsep=space]{subfig}
\usepackage{mathrsfs}
\usepackage{graphicx,float,wrapfig,epstopdf,amsmath}
\usepackage[]{algorithmicx}
\usepackage{algpseudocode,algorithm}
\usepackage{balance}

\usepackage{amsmath,mathtools }
\usepackage{amssymb}
\usepackage{amsthm}
\usepackage{graphicx}
\usepackage{epstopdf}
\usepackage{times}
\usepackage{textcomp,cite}
\usepackage{color}
\usepackage{url}
\usepackage{cuted}

\usepackage{multirow}
\usepackage[table]{xcolor}
\usepackage{threeparttable}

\begin{document}
	\title{Deep Channel Learning For Large Intelligent Surfaces Aided mm-Wave Massive MIMO Systems
		\thanks{This work was supported in part by the  ERC project  AGNOSTIC.}
	}
	\author{Ahmet~M.~Elbir\textit{, Senior Member, IEEE}, Anastasios Papazafeiropoulos\textit{, Senior Member, IEEE}, Pandelis Kourtessis, and Symeon Chatzinotas\textit{, Senior Member, IEEE}
		\thanks{A. M. Elbir is with the EE department of Duzce University, Duzce, Turkey. 
			E-mail: ahmetmelbir@gmail.com.}
		\thanks{A. Papazafeiropoulos is with the CIS Research Group, University of Hertfordshire, Hatfield, U. K. and with SnT at the University of Luxembourg, Luxembourg. E-mail: tapapazaf@gmail.com. }
		
		\thanks{P. Kourtessis is with the CIS Research Group, University of Hertfordshire, Hatfield, U. K. E-mail: p.kourtessis@herts.ac.uk}
		\thanks{S. Chatzinotas is with the SnT at the University of Luxembourg, Luxembourg. Email:symeon.chatzinotas@uni.lu. }

	}

	\maketitle
	
	\begin{abstract}
		This letter presents the first work introducing a deep learning (DL) framework for channel estimation in large intelligent surface (LIS) assisted massive MIMO (multiple-input multiple-output) systems. A twin convolutional neural network (CNN) architecture is designed and it is fed with the received pilot signals to estimate both direct and cascaded channels. In a multi-user scenario, each user has access to the CNN to estimate its own channel. The performance of the proposed DL approach is evaluated and compared with state-of-the-art DL-based techniques and its superior performance is demonstrated.
		
	\end{abstract}

	\section{Introduction}
	\label{sec:Introduciton}
	The massive MIMO (multiple-input multiple-output) architecture has been suggested as a promising technology for  fifth generation (5G) communications systems by providing high spectral efficiency  exploiting  high spatial multiplexing gains~\cite{Andrews2015,lis_SWIPT}. However, its proposed marriage with millimeter wave (mm-Wave) transmission comes with the expensive cost of energy consumption and hardware complexity, even if hybrid beamforming is employed~\cite{Andrews2015,mimoRHeath}. Recently, large intelligent surface (LIS) (also known as reflective intelligent surface) technology has been proposed as a promising solution with low cost and hardware complexity~\cite{lis_COMmag}. An LIS is an electromagnetic 2-D surface that is composed of large number of passive reconfigurable reflecting elements which are fabricated from meta-materials~\cite{lis_2018_TWC}.
	
	LIS includes a programmable meta-surface which can be controlled via external signals such as backhaul control link from the base station (BS). Hence, real-time manipulation of the reflected phase and magnitude becomes possible. This property allows us to use LIS in wireless communications as a reflecting surface between the BS and the users to improve the received signal energy, expanding the coverage as well as reducing the interference~\cite{tang2019mimo}. While LIS can provide low-cost and simplistic architecture, it brings a difficulty of including two wireless channels between the BS and user, one being the direct channel and another one is the cascaded channel between the BS and the users through LIS~\cite{lis_channelEstimation_WCL,lis_channelEst_2,lis_channelEst_3}.

	Regarding channel estimation in LIS, a transmission protocol has been proposed in \cite{lis_channelEstimation_WCL} for orthogonal frequency division multiplexing, while  \cite{lis_channelEst_2} proposed a sparse matrix factorization approach. Moreover, a dual	ascent-based estimation has been considered in \cite{lis_channelEst_3}.	 {\color{black} One of the main challenges in LIS-assisted wireless networks is that the channel estimation complexity is high due to the large number of LIS elements. To lower the complexity, deep learning (DL) techniques can be of help~\cite{lis_channelEstimation_WCL,lis_channelEst_2,lis_channelEst_3}. By training a DL network with different channel characteristics, it can adapt to the changes in the environment such as the user motions and provide robust performance. Also, updating the channel information can be done less frequently, which lowers the complexity~\cite{deepCNN_ChannelEstimation}.}	{\color{black}Notably, for LIS-assisted massive MIMO, DL has been applied for the reflected beamformer design \cite{lis_channelEst_reflectedBFDesign} and signal detection~\cite{lis_DL_detection}.  Especially, in \cite{lis_DL_detection}, the transmitted symbols are estimated without channel estimation.	This approach is symbol-dependent, i.e., when the modulation type is changed, the deep network cannot identify the symbols and requires further training.} To the best of our knowledge, this is the first work studying DL for channel estimation in an LIS scenario.
	
	In this letter, we propose a DL approach for channel estimation in a LIS-assisted mm-Wave massive MIMO systems. In the proposed DL framework, we design a twin convolutional neural network (CNN) for the estimation of direct (BS-user) and cascaded (BS-LIS-user) channels and assume that each user has access to the deep network to estimate its own channel.  The CNN is fed with the received pilot signals and it constructs a non-linear relationship between the received signals and the channel data.

	The deep network is trained with several channel realizations to obtain a robust estimation performance. In the prediction stage, a test data, which is separately generated than the training data, is used to validate the performance. Finally, we  show that the proposed DL framework achieves reasonable channel estimation accuracy and outperforms the existing DL-based techniques~\cite{deepCNN_ChannelEstimation,deepLearningChannelAndDOAEstHuang}. {\color{black}Furthermore, the results show that the proposed DL approach has robust channel estimation performance, which is tolerant to the changes in the user locations up to $4$ degrees.
	}

	\begin{figure}[h]
		\centering
		{\includegraphics[draft=false,width=.3\textheight,height=.12\textheight]{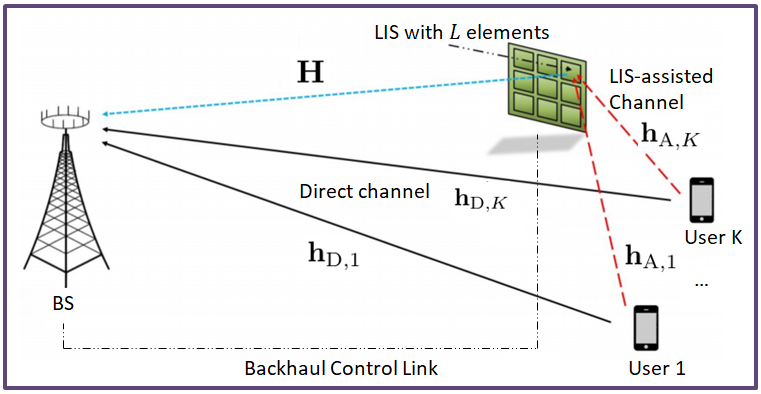} }
		\caption{An LIS-assisted mm-Wave massive MIMO scenario.}
		\label{fig_SystemArchitecture}
	\end{figure}

	\section{System Model and Problem Formulation}
	\label{sec:SystemModel}
	We consider an LIS-aided mm-Wave massive MIMO system as shown in Fig.~\ref{fig_SystemArchitecture}. We assume that the BS has $M$ antennas to serve $K$ single-antenna users with the assistance of LIS which is composed of $L$ passive reflecting elements. In LIS-assisted communication scheme, each LIS element introduces a phase shift onto the incoming signal from the BS. The phase of each LIS element can be adjusted through the PIN diodes which are controlled by the LIS-controller connected to the BS over the backhaul link~\cite{lis_channelEst_3,lis_channelEst_4}. 
	
	The BS  transmits $K$ data symbols $s_k\in\mathbb{C}$ by using a baseband precoder $\mathbf{F}= [\mathbf{f}_1,\dots,\mathbf{f}_K]\in\mathbb{C}^{M\times K}$ . Hence, the downlink $M\times 1$ transmitted  signal becomes $	\overline{\mathbf{s}} = \sum_{k=1}^{K} \sqrt{\gamma_k} \bar{\mathbf{f}}_ks_k,$
	where {\color{black} $\bar{\mathbf{f}}_k = \frac{\mathbf{f}_k}{||\mathbf{f}_k||_2}$ and} $\gamma_k$ denotes the allocated power at the $k$-th user. The transmitted signal is received from the $k$-user with two components, one of which is through the direct path from the BS and the another one is through the LIS. The received signal from the $k$-th user can be given by
	\begin{align}
	y_k = \big(\mathbf{h}_{\mathrm{D},k}^\textsf{H} + \mathbf{h}_{\mathrm{A},k}^\textsf{H} \boldsymbol{\Psi}^\textsf{H} \mathbf{H}^\textsf{H}  \big) \overline{\mathbf{s}} + n_k,
	\end{align}
	where $n_k\sim \mathcal{CN}(o,\sigma_n^2)$ and $\mathbf{h}_{\mathrm{D},k}\in\mathbb{C}^{M}$ denotes the direct channel between the BS and the $k$-th user. The vector $\mathbf{h}_{\mathrm{A},k}\in \mathbb{C}^L$ expresses the LIS-assisted channel between the LIS and the $k$-th user.  {\color{black}$\boldsymbol{\Psi}\in\mathbb{C}^{L\times L}$ is a diagonal matrix, i.e., $\boldsymbol{\Psi} = \mathrm{diag}\{\beta_1\exp(j \phi_1),\dots,\beta_L\exp(j \phi_L) \}$. Here, $\beta_l \in \{0,1\}$ represents the on/off state of the LIS elements. In practice, the LIS elements cannot be  perfectly turned on/off, Hence, $\beta_l$ can be modeled as $\beta_l=\left\{\begin{array}{cc}
		1 - \epsilon_1 & \mathrm{ON}\\
		0 + \epsilon_0 & \mathrm{OFF}
		\end{array}\right.$ for $\epsilon_1,\epsilon_0 \geq0$~\cite{lis_onoff_ICASSP}. $\phi_l\in [0,2\pi)$ is the phase shift of the reflective elements.} Finally, the channel between the LIS and the BS is represented by $\mathbf{H}\in \mathbb{C}^{M\times L}$.

	%
	%
	%
	
	In mm-Wave transmission, the channel can be represented by the Saleh-Valenzuela (SV) model where a geometric channel model is adopted with limited scattering~\cite{lis_channelEst_4,lis_channelEstimation_WCL}. Hence, we assume that the mm-Wave channels, i.e., $\mathbf{h}_{\mathrm{D},k}, \mathbf{h}_{\mathrm{A},k}$ and $\mathbf{H}$, include the contributions of $N_\mathrm{D}$, $N_\mathrm{A}$ and $N_\mathrm{H}$ paths, respectively. Thus, we can represent the channels $\mathbf{h}_{\mathrm{D},k}$ and $\mathbf{h}_{\mathrm{A},k}$ as
	$\mathbf{h}_{\mathrm{D},k} = \sqrt{\frac{M }{N_\mathrm{D}  }} \sum_{n_\mathrm{D}=1}^{ N_\mathrm{D}} \alpha_{\mathrm{D},k}^{(n_\mathrm{D})} \mathbf{a}_\mathrm{D}( \theta_{\mathrm{D},k}^{(n_\mathrm{D})}),$ and $\mathbf{h}_{\mathrm{A},k} = \sqrt{\frac{L }{N_\mathrm{A}  }} \sum_{n_\mathrm{A}=1}^{ N_\mathrm{A}} \alpha_{\mathrm{A},k}^{(n_\mathrm{A})} \mathbf{a}_\mathrm{A}( \theta_{\mathrm{A},k}^{(n_\mathrm{A})}),$
	where $\{\alpha_{\mathrm{D},k}^{(n_\mathrm{D})},\alpha_{\mathrm{A},k}^{(n_\mathrm{A})}\}$ and $\{\theta_{\mathrm{D},k}^{(n_\mathrm{D})},\theta_{\mathrm{A},k}^{(n_\mathrm{A})}  \}$ are the complex channel gains and received path angles for the corresponding channels, respectively. $\mathbf{a}_\mathrm{D}( \theta)$ and $\mathbf{a}_\mathrm{A}( \theta)$ are $M\times 1$ and $L\times 1$ steering vectors of the path angles as $\mathbf{a}_\mathrm{D}( \theta) = \frac{1}{\sqrt{M}}[e^{j \omega_0},\dots, e^{j \omega_{M-1}}]^\textsf{T}$, $\mathbf{a}_\mathrm{A}( \theta) = \frac{1}{\sqrt{L}}[e^{j\omega_0},\dots,e^{j \omega_{L-1}}]^\textsf{T}$ where $\omega_{n} = n\frac{2\pi d}{\lambda}\pi\sin(\theta)$  and $d = \lambda/2$ is the array spacing for the wavelength $\lambda$. Further, the mm-Wave channel between the BS and the LIS is given by
	\begin{align}
	\label{eq:ChannelModel}
	\mathbf{H} =  \sqrt{\frac{ M L } {N_\mathrm{H} }}\sum_{n_\mathrm{H}=1}^{N_\mathrm{H}}    \alpha^{(n_\mathrm{H})} \mathbf{a}_\mathrm{BS}( \theta_{\mathrm{BS}}^{(n_\mathrm{H})}) \mathbf{a}_\mathrm{LIS}^\textsf{H}( \theta_{\mathrm{LIS}}^{(n_\mathrm{H})}),
	\end{align}  
	where $\alpha^{(n_\mathrm{H})}\in \mathbb{C}$ denotes the complex gain and $\{\theta_{\mathrm{BS}}^{(n_\mathrm{H})},\theta_{\mathrm{LIS}}^{(n_\mathrm{H})}\}$ are the angle-of-departure (AOD) and angle-of-arrival (AOA) angles of the paths, respectively. $\mathbf{a}_\mathrm{BS}( \theta)\in\mathbb{C}^{M}$ and $\mathbf{a}_\mathrm{LIS}( \theta)\in\mathbb{C}^L$ are the steering vectors.	Let $\mathbf{G}_{k}\in \mathbb{C}^{M\times L}$ be the cascaded channel matrix between the BS and the $k$-th user as $\mathbf{G}_k = \mathbf{H}\boldsymbol{\Gamma}_k$ where $\boldsymbol{\Gamma}_k = \mathrm{diag}\{ \mathbf{h}_{\mathrm{A},k}\}$. Then, we can write $\mathbf{H} \boldsymbol{\Psi}\mathbf{h}_{\mathrm{A},k} = \mathbf{G}_k \boldsymbol{\psi}$, for which we have $\boldsymbol{\Psi} = \mathrm{diag}\{\boldsymbol{\psi} \}$.

	In this work, our aim is to  estimate the direct  and cascaded channels $\{\mathbf{h}_{\mathrm{D},k}, \mathbf{G}_k \}$ in downlink transmission. In this case, we assume that each user feeds the received pilot signals to the deep network (henceforth, called \textsf{ChannelNet}) to estimate its own channel.
	
	

	\section{Channel Estimation Via Deep Learning}
	\label{sec:DL}
	The proposed DL framework uses the received pilot signals as input to estimate the direct and cascaded channels.
	
	\subsection{Labeling}
	Consider the downlink scenario where the BS transmits the orthogonal pilot signals $\mathbf{x}_p\in\mathbb{C}^M$, one at a single coherence time $\tau$, with $p= 1,\dots,P$ and $P\geq M$.  Hence, the total number of  channel uses to estimate the direct channel is $P$ The received signal at the $k$-th user can be given by
	\begin{align}
	\label{receivedDirectChannel}
	\mathbf{y}_{k} = \big(\mathbf{h}_{\mathrm{D},k}^\textsf{H} + \boldsymbol{\psi}^\textsf{H}\mathbf{G}_k^\textsf{H}  \big)\mathbf{X} + \mathbf{n}_{k},
	\end{align}
	where $\mathbf{X} = [\mathbf{x}_1,\dots, \mathbf{x}_P]\in\mathbb{C}^{M\times P}$ is the pilot signal matrix while $\mathbf{y}_k = [{y}_{k,1},\dots, {y}_{k,P}]$ and $\mathbf{n}_k = [{n}_{k,1},\dots, {n}_{k,P}]$ are ${1\times P}$ row vectors and  $\mathbf{n}_k\sim \mathcal{CN}(0,\sigma_n^2\boldsymbol{\mathrm{I}}_{{P}})$.  We assume that the pilot training has two phases: direct channel estimation (i.e., $\mathbf{h}_{\mathrm{D},k}$) and the cascaded channel estimation (i.e., $\mathbf{G}_k$). In phase I, we assume that all of the LIS elements are turned off, i.e.,  $\beta_l = 0, \forall l$, by using the BS backhaul link{\color{black}\footnote{\color{black} When the PIN diodes are turned off, the reflecting elements are almost transparent with the insertion loss nearly being  zero such that the incoming signals pass through the reflecting elements~\cite{lis_diode_experiment}.}}. {\color{black}We note here that by setting $\beta_l$ as $\{1,0\}$ does not affect the the direct and cascaded channels since they do not depend on the reflect beamformer $\boldsymbol{\Psi}$ as seen in (\ref{receivedDirectChannel}).} Then, the received baseband signal at the $k$-th user becomes
	\begin{align}
	\label{receivedPilot_DC}
	\mathbf{y}_{\mathrm{D}}^{(k)} =\mathbf{h}_{\mathrm{D},k}^\textsf{H} \mathbf{X} + \mathbf{n}_{\mathrm{D},k}.
	\end{align}
	Here, the direct channel $\mathbf{h}_{\mathrm{D},k}$ is selected as the label of the deep network with the corresponding input data of  $	\mathbf{y}_{\mathrm{D}}^{(k)}$. 
	
	Once ${\mathbf{h}}_{\mathrm{D},k}$, being the estimated channel, is obtained, in the second phase of the training stage, the cascaded channel $\mathbf{G}_k$ can be estimated. {\color{black}This can be achieved via two approaches. In the first approach, $P=M$ pilot signals are transmitted} when each of the LIS elements is turned on one by one.  In this case, the BS sends a request to LIS via the micro-controller device in the backhaul link to turn on a single LIS element at a time.	For the $l$-th frame, the reflect beamforming vector becomes $\boldsymbol{\psi}^{(l)} = [0,\dots, 0, \psi_l, 0, \dots, 0]^\textsf{T}$ where {\color{black}$\beta_{\bar{l}} =\{ 0:\bar{l}=1,\dots,L, \bar{l} \neq l\}$} and the received signal from the cascaded channel at the $k$-th user becomes
	\begin{align}
	\label{receivedPilot_CC}
	\mathbf{y}_\mathrm{C}^{(k,l)} =  \big(\mathbf{h}_{\mathrm{D},k}^\textsf{H} + \mathbf{g}_{k,l}^\textsf{H}  \big) \mathbf{X} + \mathbf{n}_{k,l},
	\end{align}
	where $\mathbf{y}_\mathrm{C}^{(k,l)}  =[y_{\mathrm{C},1}^{(k,l)},\dots, y_{\mathrm{C},P}^{(k,l)}]$ and $\mathbf{n}_{k,l} =[n_{k,1}^{(l)},\dots, n_{k,P}^{(l)}]$ are $1\times P$ row vectors. In (\ref{receivedPilot_CC}), $\mathbf{g}_{k,l}$ represents the $l$-th column of $\mathbf{G}_k$ as $\mathbf{g}_{k,l}=\mathbf{G}_{k}\boldsymbol{\psi}^{(l)}$. 
	Then the least-squares (LS) estimate of $\mathbf{g}_{k,l}$ becomes
	\begin{align}
	\label{cascadedChannelEst}
	\hat{\mathbf{g}}_{k,l} =  \big(\mathbf{y}_\mathrm{C}^{(k,l)}\mathbf{X}^\textsf{H} \big(\mathbf{X} \mathbf{X}^\textsf{H} \big)^{-1}\big)^\textsf{H}   -  \mathbf{h}_{\mathrm{D},k}.
	\end{align}
	By using $\hat{\mathbf{h}}_{\mathrm{D},k}$, (\ref{cascadedChannelEst}) can be solved for $l = 1\dots,L$. Then, we can construct the estimated cascaded matrix as $\hat{\mathbf{G}}_k = [\hat{\mathbf{g}}_{k,1},\dots, \hat{\mathbf{g}}_{k,L}]$.
	
	{\color{black}In the second approach, channel estimation is done when all LIS elements are turned on. In this case, the $L$ columns of $\mathbf{G}_k$ are jointly estimated by using $\bar{\mathbf{X}}\in\mathbb{C}^{ML\times ML}$ pilot signal matrix. Let the reflect beamforming vector be an $L\times 1$ vector of all ones, i.e., $\bar{\boldsymbol{\psi}}=\mathbf{1}_L$, then we can write the $ML\times 1$ received signal as
		\begin{align}
		\label{outputCascade2}
		\bar{\mathbf{y}}_\mathrm{C}^{(k)} = \big( \bar{\mathbf{h}}_{\mathrm{D},k}^\textsf{H}  + \bar{\mathbf{g}}_k^\textsf{H}  \big)  \bar{\mathbf{X}} + \bar{\mathbf{n}}_k,
		\end{align}
		where $\bar{\mathbf{h}}_{\mathrm{D},k} = \mathbf{1}_L\otimes {\mathbf{h}}_{\mathrm{D},k}$ where $\otimes$ denotes the kronecker product  and  $\bar{\mathbf{g}}_k = [{\mathbf{g}}_{k,1}^\textsf{T},\dots, {\mathbf{g}}_{k,L}^\textsf{T}]^\textsf{T}$. Then, the LS estimate of $\bar{\mathbf{g}}_k$ becomes
		\begin{align}
		\label{cascadedChannelEst2}
		\hat{\bar{\mathbf{g}}}_k = \big(\bar{\mathbf{y}}_\mathrm{C}^{(k)}\bar{\mathbf{X}}^\textsf{H} \big(\bar{\mathbf{X}} \bar{\mathbf{X}}^\textsf{H} \big)^{-1}\big)^\textsf{H}   -  \bar{\mathbf{h}}_{\mathrm{D},k}.
		\end{align}
		
		The estimated cascaded channel from (\ref{cascadedChannelEst}) and (\ref{cascadedChannelEst2}) will yield the same results if perfectly orthogonal pilots are used. When the pilot signals become correlated/corrupted, then (\ref{cascadedChannelEst}) provides better results since $\mathbf{X}$ involves less corruption than $\bar{\mathbf{X}}$ (Please see Fig.~\ref{fig_SNR_X_Rate}).
	}


	%
	%


	\begin{algorithm}[h]
		\begin{algorithmic}[1]
			\caption{Training data generation for \textsf{ChannelNet}. }
			\Statex {\textbf{Input:} $K$,  $U$, $V$,  $\mathbf{X}$, $\boldsymbol{\psi}$ SNR,  SNR$_{{\mathbf{h}}}$, SNR$_\mathbf{G}$}. \\
			{\textbf{Output:} Training datasets $\mathcal{D}_\mathrm{DC}$ and $\mathcal{D}_\mathrm{CC}$.}
			\label{alg:algorithmTraining}
			\State Initialize with $t=1$ and the dataset length is $T=UVK$.
			\State   \textbf{for}  $1 \leq v \leq V$ \textbf{do}
			\State \indent Generate $\mathbf{h}_{\mathrm{D},k}^{(v)}$ and $\mathbf{G}_k^{(v)}$ {\color{black} from Section~\ref{sec:SystemModel}}
			, $\forall k$.
			\State  \indent \textbf{for}  $1 \leq u \leq U$ \textbf{do}
			\State \indent  $[{\mathbf{h}}_{\mathrm{D},k}^{(u,v)}]_{i,j} \sim \mathcal{CN}([\mathbf{h}_{\mathrm{D},k}^{(v)}]_{i,j},\sigma_{\mathbf{h}}^2)$, $\forall k$.
			\State \indent  $[{\mathbf{G}}_k^{(u,v)}]_{i,j} \sim \mathcal{CN}([\mathbf{G}_k^{(v)}]_{i,j},\sigma_{\mathbf{G}}^2)$, $\forall k$.
			\State \indent \textbf{for} $1\leq k \leq K$ \textbf{do} 
			\State \indent Using \hspace{-2pt}$\mathbf{h}_{\mathrm{D},k}^{(u,v)}$\hspace{-3pt} and $\mathbf{g}_{k,l}^{(u,v)}$, \hspace{-3pt} generate \hspace{-3pt} $\mathbf{y}_{\mathrm{D}}^{{(k)}^{(u,v)}}$ \hspace{-5pt}and \hspace{-3pt} $\mathbf{y}_\mathrm{C}^{{(k,l)}^{(u,v)}}$ \hspace{-3pt} \par \indent from (\ref{receivedPilot_DC}) an (\ref{receivedPilot_CC}).
			\State \indent Using\hspace{-1pt} $\mathbf{y}_{\mathrm{D}}^{{(k)}^{(u,v)}}$ \hspace{-5pt}and \hspace{-3pt} $\mathbf{y}_\mathrm{C}^{{(k,l)}^{(u,v)}}$ \hspace{-11pt}; design $\mathbf{X}_{\mathrm{DC}}^{(t)}$ and  $\mathbf{X}_{\mathrm{CC}}^{(t)}$.
			\State \indent  Using $\mathbf{h}_{\mathrm{D},k}^{(u,v)}$, $\mathbf{G}_{k}^{(u,v)}$; design the output $\mathbf{z}_{\mathrm{DC}}^{(t)}$, $\mathbf{z}_\mathrm{CC}^{(t)}$.
			\State \indent $\mathcal{D}_\mathrm{DC}^{(t)} = (\mathbf{X}_\mathrm{DC}^{(t)}, \mathbf{z}_\mathrm{DC}^{(t)})$,  $\mathcal{D}_\mathrm{CC}^{(t)} = (\mathbf{X}_\mathrm{CC}^{(t)}, \mathbf{z}_\mathrm{CC}^{(t)})$.
			\State \indent $t \leftarrow t + 1,$
			\State \indent \textbf{end for} $k$,
			\State \indent \textbf{end for} $u$,
			\State \textbf{end for} $v$,
			
		\end{algorithmic}
	\end{algorithm}

	\begin{figure}[]
		\centering
		{\includegraphics[draft=false,width=.35\textheight,height=.12\textwidth]{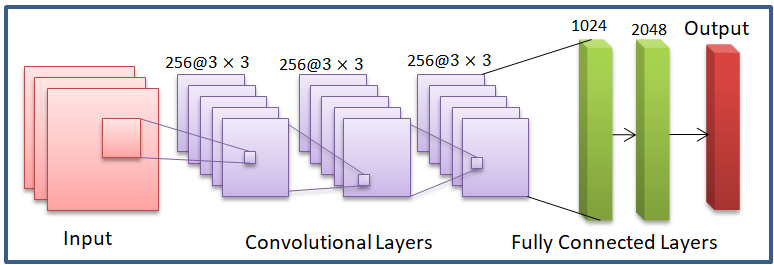} } 
		\caption{Proposed deep neural network architecture.   }
		\label{fig_deepNet}
	\end{figure}
	

	\subsection{Input Design: Received Pilots}
	\label{sec:input}
	The deep network accepts the received signals as input  at the preamble stage. As a result, the input-output pairs become $\{\mathbf{y}_{\mathrm{D}}^{(k)}, \mathbf{h}_{\mathrm{D},k}\}$ and $\{\mathbf{y}_\mathrm{C}^{(k,l)},\mathbf{g}_{k,l}\}$ for direct and cascaded channel estimation, respectively.
	In order to feed the deep network we use real, imaginary and the absolute value of each entry of the received signal. While the use of only real/imaginary components is still possible~\cite{deepCNN_ChannelEstimation}, it is shown in~\cite{elbirIETRSN2019,elbirDL_COMML,elbirQuantized_TWC_2020,elbirTVT_2020} that the use of ``three-channel" data ameliorates the performance by enriching the features inherited in the input data. Let us define the input of the deep network as $\mathbf{X}_\mathrm{DC}$ and $\mathbf{X}_{\mathrm{CC}}$ for the direct and cascaded channel, respectively. Then, $\mathbf{X}_\mathrm{DC}$ is a $\tilde{M}\times \tilde{M}\times 3$ real-valued ``three-channel" matrix, each of which is of size $\tilde{M}=\sqrt{M}$. In order to benefit from  2-D convolutional filters, we construct the matrix quantity $\mathbf{X}_{\mathrm{DC}}$  from the vector $\mathbf{y}_{\mathrm{D}}^{(k)}$ by partitioning $\mathbf{y}_{\mathrm{D},k}$ into $\sqrt{M}$ subvectors and put them into $\sqrt{M}$ columns\footnote{ $\sqrt{M}$ is assumed to be an integer value. If not, a rectangular $\mathbf{X}_\mathrm{DC}$ can always be constructed without affecting the network training.}. In particular, for the first and the second ``channels" of $\mathbf{X}_{\mathrm{DC}}$, we have $\mathrm{vec} \big\{[\mathbf{X}_{\mathrm{DC}}]_1 \big\} = \operatorname{Re}\big\{ \mathbf{y}_{\mathrm{D}}^{(k)} \big\}$  and $\mathrm{vec} \big\{[\mathbf{X}_{\mathrm{DC}}]_2 \big\} = \operatorname{Im}\big\{ \mathbf{y}_{\mathrm{D}}^{(k)} \big\}$. Finally, the third ``channel" is denoted by the element-wise absolute value of $\mathbf{y}_{\mathrm{D}}^{(k)}$ as  $\mathrm{vec} \big\{[\mathbf{X}_{\mathrm{DC}}]_3 \big\} = \big| \mathbf{y}_{\mathrm{D}}^{(k)} \big|$. Similarly, we can define $\mathbf{X}_\mathrm{CC}$ as an $L\times M\times 3$ real-valued matrix and we have {\color{black} $\mathrm{vec} \big\{[\mathbf{X}_{\mathrm{CC}}]_1 \big\} = \operatorname{Re}\big\{ \tilde{\mathbf{y}}_{k} \big\}$, $\mathrm{vec} \big\{[\mathbf{X}_{\mathrm{CC}}]_2 \big\} = \operatorname{Im}\big\{ \tilde{\mathbf{y}}_{k} \big\}$ and  $\mathrm{vec} \big\{[\mathbf{X}_{\mathrm{CC}}]_3 \big\} = \big| \tilde{\mathbf{y}}_{k} \big|$ where $\tilde{\mathbf{y}}_{k} = [\mathbf{y}_\mathrm{C}^{(k,l)^\textsf{T}},\dots, \mathbf{y}_\mathrm{C}^{(k,L)^\textsf{T}}]^\textsf{T}$	is an $LM\times 1$ vector composed of the received pilot signals. The input design for the second approach can also be done accordingly by using (\ref{outputCascade2}).} The output of the deep network is the vectorized form of the channel matrices, i.e., $\mathbf{z}_\mathrm{DC} = \big[\operatorname{Re}\{\mathbf{h}_{\mathrm{D},k}\}^\textsf{T}, \operatorname{Im}\{\mathbf{h}_{\mathrm{D},k}\}^\textsf{T}\big]^\textsf{T} $ and $\mathbf{z}_\mathrm{CC} = \big[\operatorname{Re}\{\mathrm{vec}\{\mathbf{G}_{k}\}\}^\textsf{T}, \operatorname{Im}\{\mathrm{vec}\{\mathbf{G}_{k}\}\}^\textsf{T}\big]^\textsf{T}$ which are $2M\times 1$ and $2ML \times 1$ vectors, respectively. The training data can be obtained by generating the input-output pairs for several realizations, as described in Algorithm~\ref{alg:algorithmTraining}.

	\subsection{Network Architectures and Training}
	\label{Training}
	\textsf{ChannelNet} composed of two identical CNNs, each of which is composed of $9$ layers as illustrated in Fig.~\ref{fig_deepNet}. The first layer is the input layer which accepts the received pilot signals. Since the same network is used for both direct and cascaded channels, the size of the input differs as described in Section~\ref{sec:input}. The $\{2, 3, 4\}$-th layers are convolutional layers (CLs) with $256$ filters of size $3\times 3$. The fifth and the seventh layers are fully connected layers (FCLs) with $1024$ and $2048$ units respectively. There are dropout layers  with a $50\%$ probability after each FCL and the last layer is the regression layer. The network parameters are fixed after a hyperparameter tuning process that yields the best performance for the considered scenario \cite{elbirDL_COMML,elbirQuantized_TWC_2020,elbirIETRSN2019}. The proposed deep network is realized and trained in MATLAB on a PC with a single GPU and a 768-core processor. We used the stochastic gradient descent algorithm with momentum 0.9  and  updated the network parameters with learning rate $0.0002$ and mini-batch size of $128$ samples. Then, we  applied a stopping criterion in training which ceases  training  when the validation accuracy does not improve in three consecutive epochs.

	\begin{figure}[t]
		\centering
		\subfloat{\includegraphics[draft=false,width=.32\textheight,height=.15\textheight]{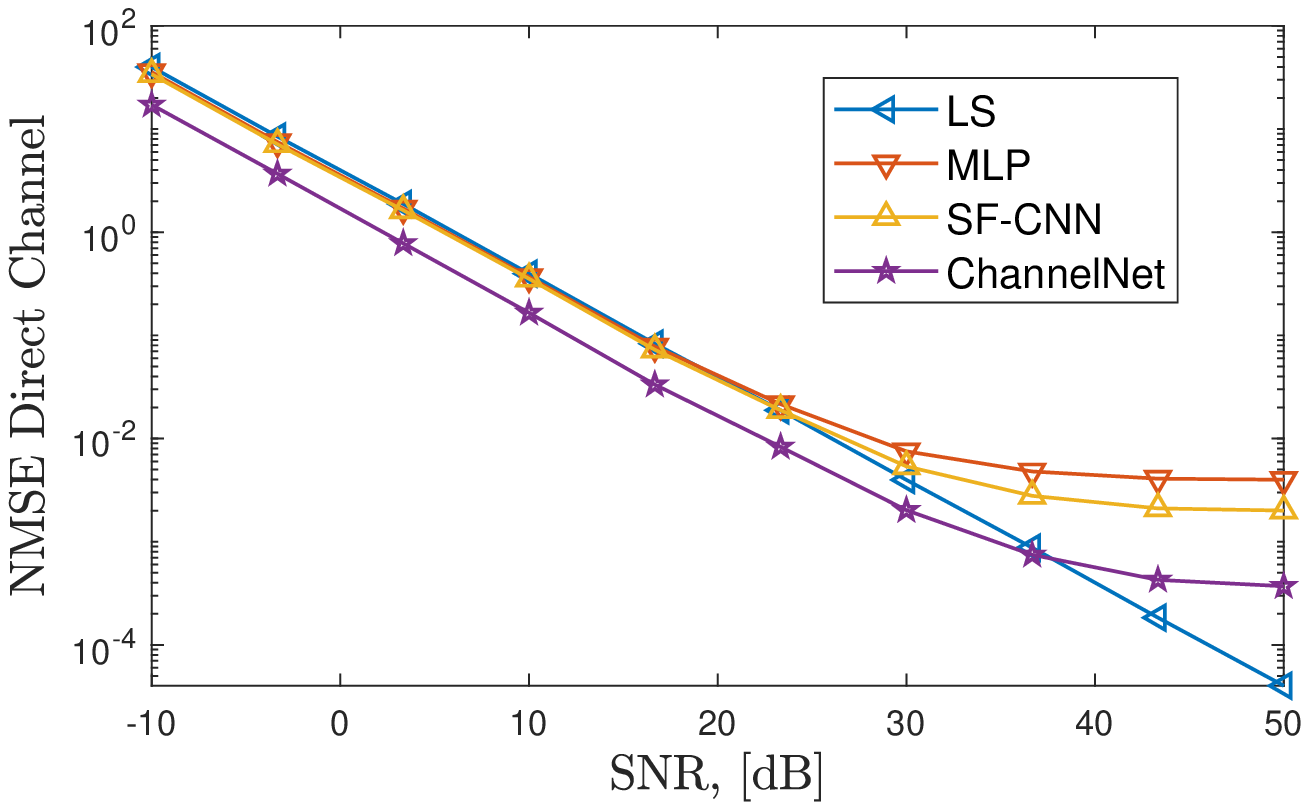} }\\ 
		\subfloat{\includegraphics[draft=false,width=.32\textheight,height=.15\textheight]{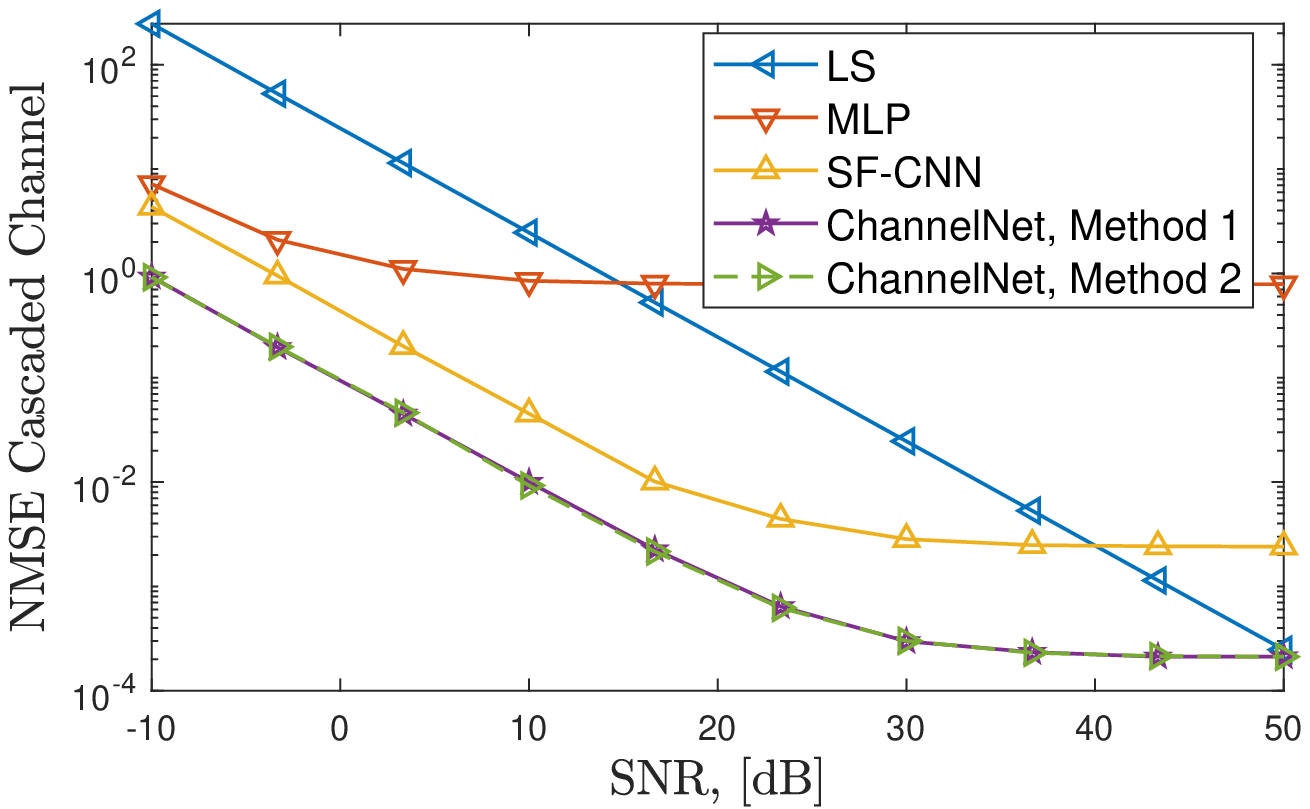} }
		\caption{Channel estimation NMSE with respect to SNR.}
		\label{fig_SNR_Rate}
	\end{figure}
	

	%
	
	\section{Numerical Simulations}
	\label{sec:Sim}
	In this part, we evaluate the performance of the proposed \textsf{ChannelNet} framework with comparison to the state-of-the-art DL-based approaches such as MLP~\cite{deepLearningChannelAndDOAEstHuang} and SF-CNN~\cite{deepCNN_ChannelEstimation}. Throughout the simulations, we select $P=M=64$, $L=100$, and $K=8$. The physical environment is modeled with $N_\mathrm{D}=N_\mathrm{A} = N_\mathrm{H}  = 10$ paths, where the direction of users are uniform randomly drawn from the interval $[-\pi,\pi]$ and we select $\epsilon_0=\epsilon_1 = 0$ unless stated otherwise.
	
	To train the network, we generate different channel scenarios for $U=100$, $V = 500$ and $K=8$. During training, three  signal-to-noise ratio
	(SNR)  levels are used to improve the robustness, i.e., SNR= $\{10,20,30\}$ dB. In addition, synthetic noise is added to the labels with SNR$_\mathbf{h} =$SNR$_\mathbf{G} =\{20,30\}$ dB, where SNR$_\mathbf{h} = 20\log_{10}(\frac{|[\mathbf{h}]_{i}|^2}{\sigma_{\mathbf{h}}^2})$ and SNR$_\mathbf{G} = 20\log_{10}(\frac{|[\mathbf{G}]_{i,j}|^2}{\sigma_{\mathbf{G}}^2})$, respectively. Hence, the total data length is  $T = 240000$. During training $70\%$ and $30\%$ of the whole generated data are used for training and validation respectively. {\color{black} The training stage takes about $40$ minutes, whereas the online deployment of the proposed DL network only needs $0.004$ seconds.} Once the training is completed, a new received pilot data other than the training data is generated and used in the prediction stage, where $J=100$ Monte Carlo experiments are conducted to assess the normalized mean-square-error (NMSE) performance i.e.,  the NMSE for $\mathbf{G}_k$ is defined as $ \frac{1}{J}\sum_{j=1}^{J}  {|| \mathbf{G}_k - \hat{\mathbf{G}}_k^{(j)} ||_\mathcal{F}}/{|| \mathbf{G}_k ||_\mathcal{F}   }$.

	\begin{figure}[h]
		\centering
		\subfloat{\includegraphics[draft=false,width=.32\textheight,height=.15\textheight]{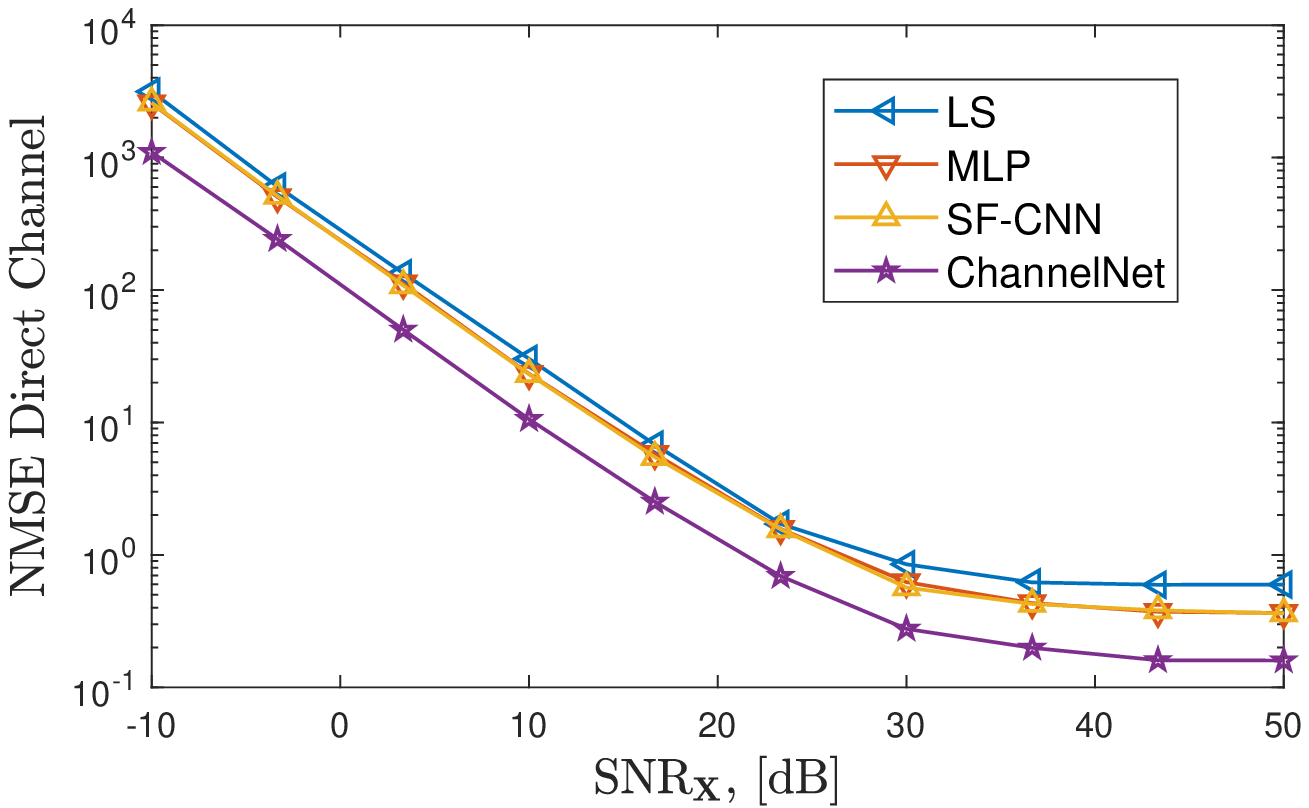} }\\ 
		\subfloat{\includegraphics[draft=false,width=.32\textheight,height=.15\textheight]{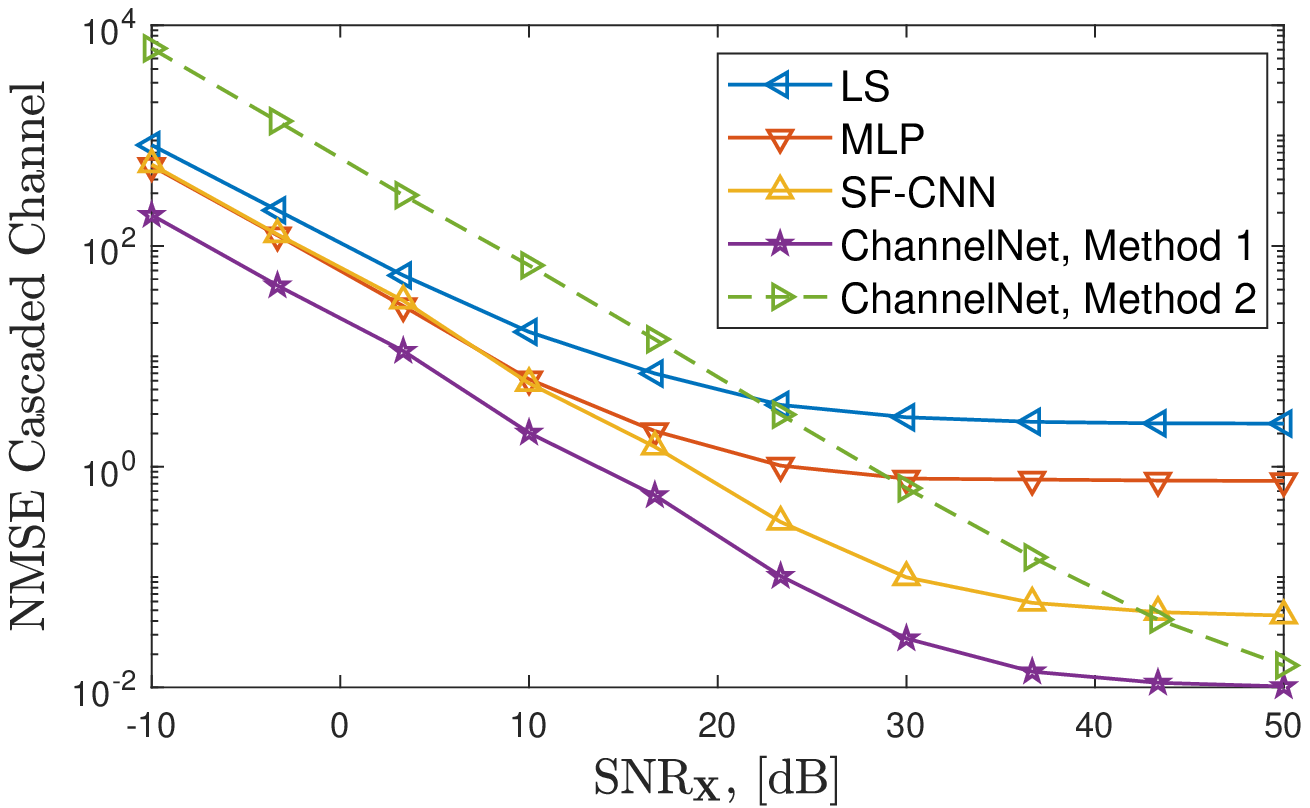} }
		\caption{Channel estimation NMSE with respect to SNR$_{\mathbf{X}}$.}
		\label{fig_SNR_X_Rate}
	\end{figure}
	

	In Fig.~\ref{fig_SNR_Rate}, we present the channel estimation  for direct and cascaded channels with respect to SNR. We can see that the DL-based approaches have better NMSE than the LS~\cite{lis_channelEstimation_WCL} due to their better mapping architectures from the received pilots to channel data. Among the DL-based techniques, \textsf{ChannelNet} has superior performance as compared to the others. The effectiveness of \textsf{ChannelNet} is due to the joint use of CL and FCLs, whereas MLP and SF-CNN have FCL-only and CL-only structures, respectively. While FCLs are powerful in constructing non-linear mapping between input and the output, CLs play very important role in DL networks when generating new features to enrich the mapping performance.  We also observe that the performance of the DL-based approaches saturates at high SNR (i.e., $>20$ dB) because of the biased nature of the neural networks which do not provide unlimited accuracy. This problem can be mitigated by increasing the number of units in various network layers. Unfortunately, it may lead to  network memorizing the training data and perform poorly when the test data are different than the ones in training. To balance this trade-off, we have used noisy data-sets during training so that the network attains reasonable tolerance to corrupted/imperfect inputs.

	In Fig.~\ref{fig_SNR_X_Rate}, the effect of corrupted pilot data is examined and the performance of the algorithms is obtained with respect to SNR on the pilot data, i.e., SNR$_\mathbf{X} =20\log_{10}(\frac{|[\mathbf{X}]_{i,j}|^2}{\sigma_{\mathbf{X}}^2}) $ when SNR$=10$ dB. We observe that all of the algorithms require at least SNR$_\mathbf{X}\geq 20$ dB to provide a reasonable NMSE performance and the proposed DL approach has the superior performance among all.{\color{black} We see that cascaded channel estimation with the first method is more robust to pilot corruption than the second method due the use of fewer pilot signals.}

	\begin{figure}[h]
		\centering
		\subfloat{\includegraphics[draft=false,width=.32\textheight,height=.15\textheight]{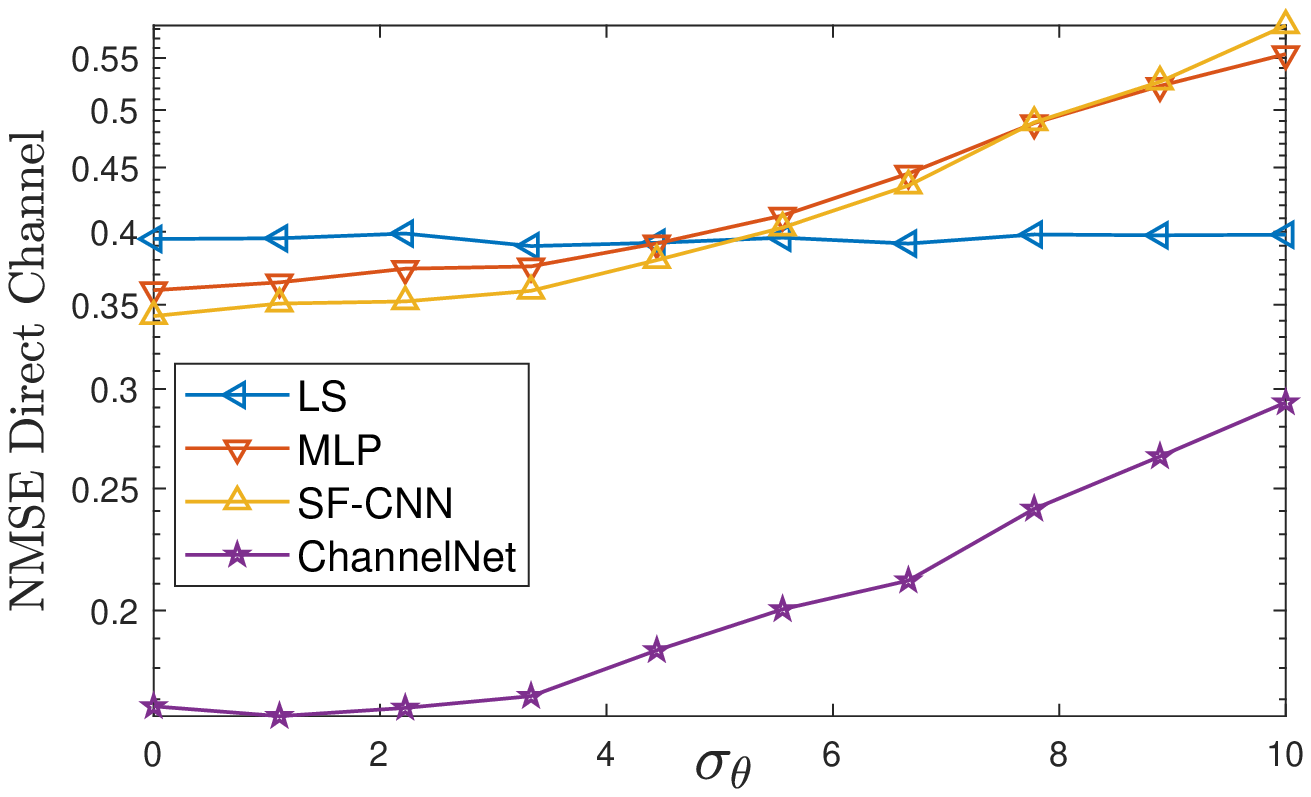} }\\ 
		\subfloat{\includegraphics[draft=false,width=.32\textheight,height=.15\textheight]{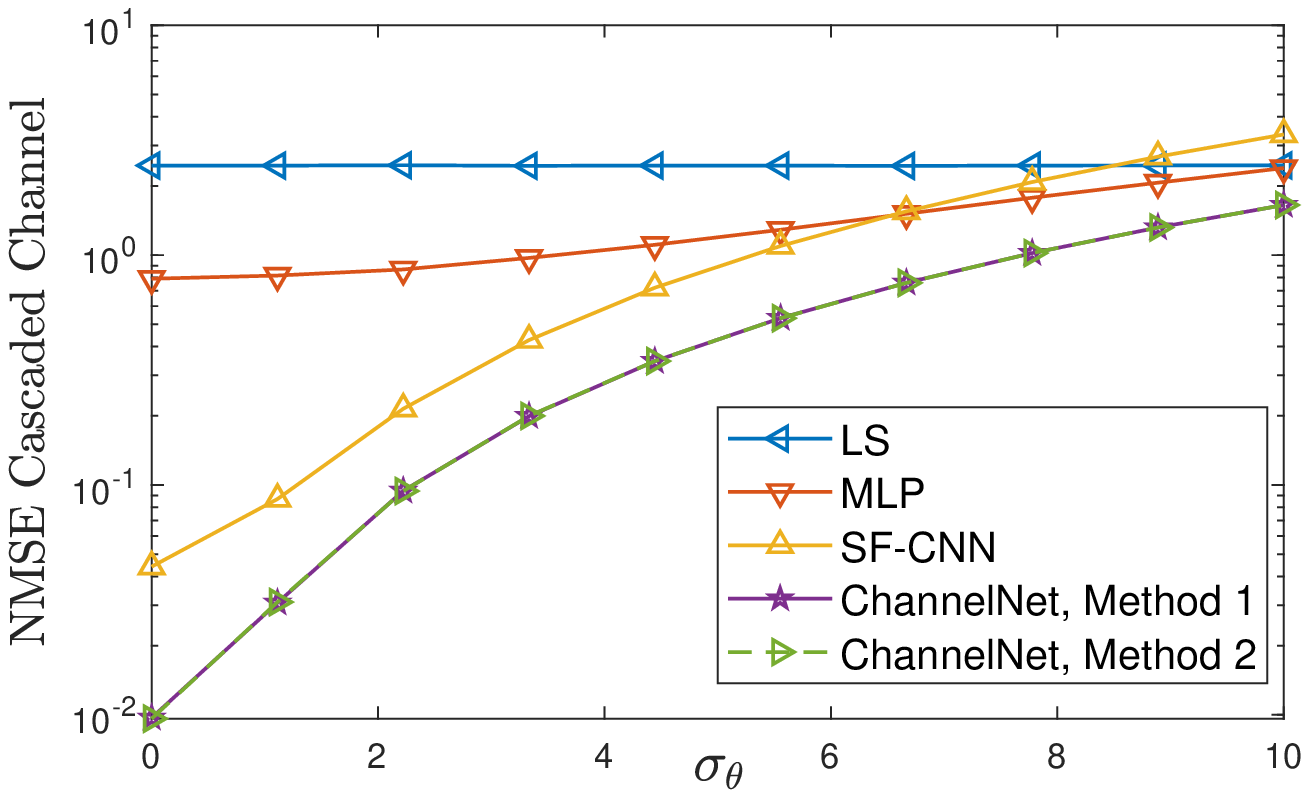} }
		\caption{Channel estimation NMSE with respect to $\sigma_\theta$.}
		\label{fig_Angle}
	\end{figure}
	
	
	\begin{figure}[h]
		\centering
		\subfloat{\includegraphics[draft=false,width=.32\textheight,height=.15\textheight]{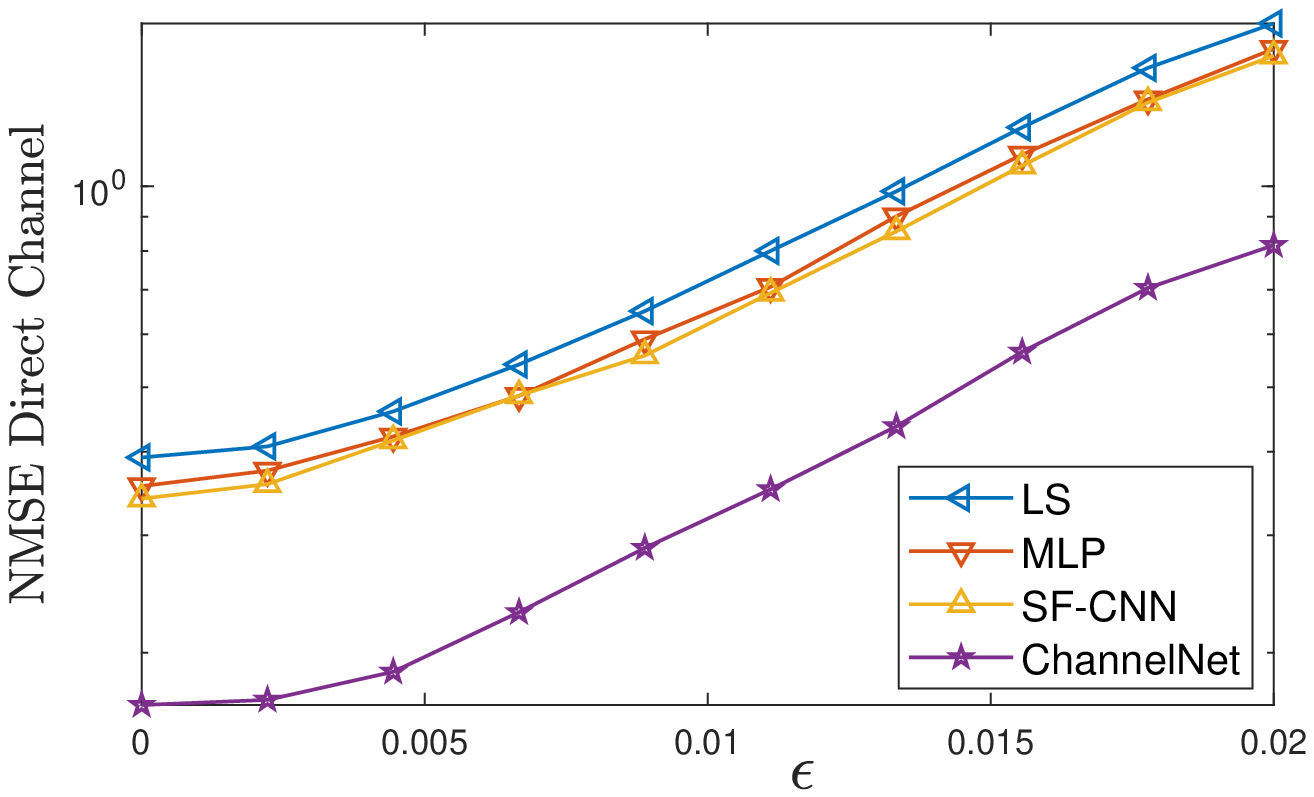} }\\ 
		\subfloat{\includegraphics[draft=false,width=.32\textheight,height=.15\textheight]{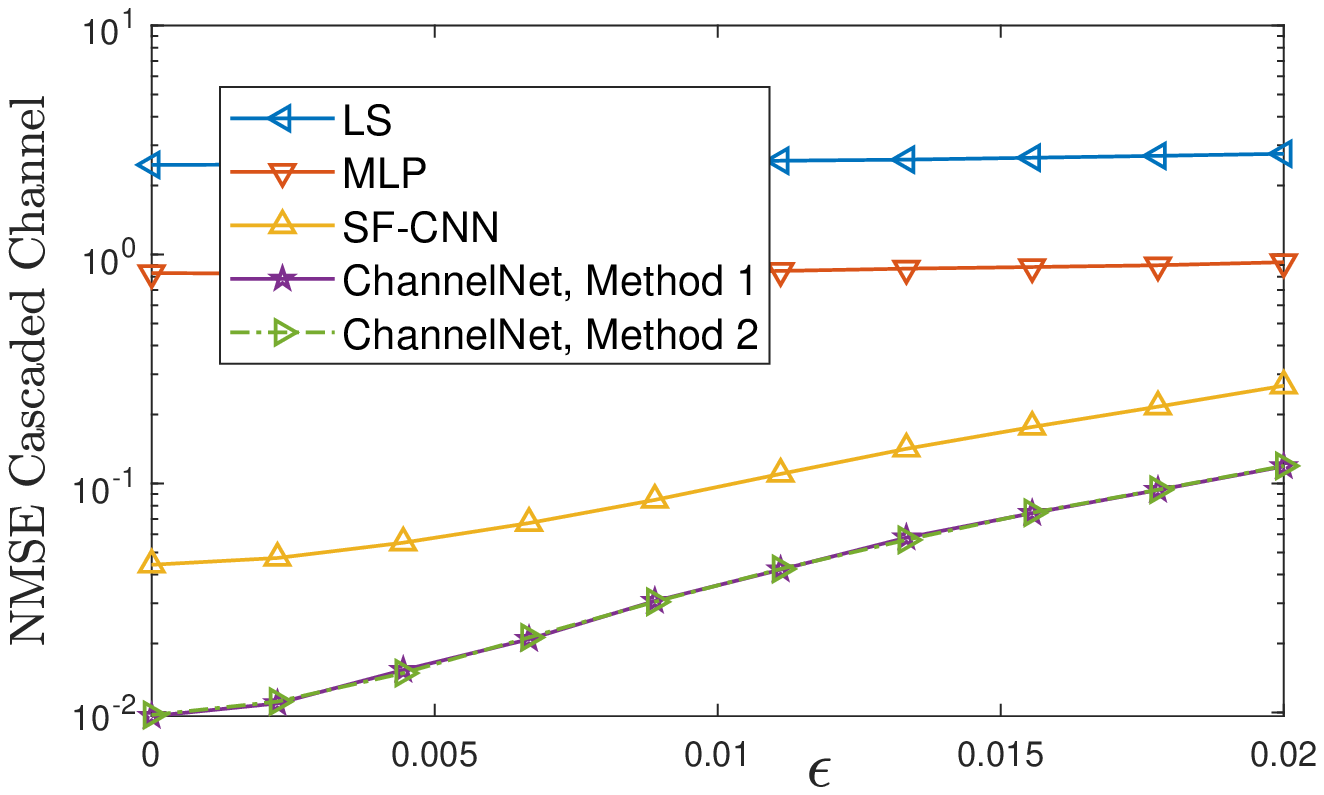} }
		\caption{Channel estimation NMSE with respect to $\epsilon$.}
		\label{fig_Eps}
	\end{figure}
	
	
	{\color{black}The adaptation of DL-based techniques when the channel data change is an important performance  measure. In Fig.~\ref{fig_Angle}, the NMSE is presented when the AOAs of the users in the test stage are different for training. In the test stage, we introduce an angle mismatch into the AOA of all users with standard deviation $\sigma_\theta$ and present the results in Fig.~\ref{fig_Angle}. We can see that \textsf{ChannelNet} outperforms the other algorithms and it provides satisfactory performance up to $4^\circ$ angular mismatch in the test data.

		In Fig.\ref{fig_Eps}, we present the performance with respect to non-ideal switching of LIS elements for $\epsilon = \epsilon_0=\epsilon_1$. As it is seen, $\epsilon \leq 5\times 10^{-3}$ provides satisfactory NMSE performance.
	}

	\section{Summary}
	\label{sec:Conc}
	We proposed a DL-based channel estimation technique for LIS-assisted massive MIMO systems. In the proposed scheme, each user has an identical deep network which is fed by the received pilot signals to  effectively estimate the direct and the cascaded channels. {\color{black} We have conducted several experiments to investigate the performance of the algorithms and observed that the proposed approach outperforms the other algorithms. We have shown that the proposed approach does not need to be re-trained when the user locations change up to 4 degrees. We have also investigated the non-ideal switching scenario for the LIS elements and shown that the proposed method can provide reasonable performance up to $0.5\%$ amplitude error in switching.}

	\bibliographystyle{IEEEtran}
	\footnotesize{\bibliography{IEEEabrv,references_054_journal}}
	\balance

\end{document}